\documentclass[10pt]{iopart}

\usepackage{iopams}
\usepackage[T1]{fontenc}
\usepackage[utf8]{inputenc}
\usepackage{graphicx}
\usepackage{xcolor}
\usepackage{textcomp}
\usepackage{enumerate}
\usepackage{todonotes}
\usepackage{hyperref}
\usepackage{comment}
\usepackage{url}
\usepackage{iopams}
\usepackage{cite}
\usepackage{url}
\usepackage{ulem}
\usepackage{soul}
\usepackage{dcolumn}
\newcolumntype{C}[1]{>{\centering}p{#1}}
\usepackage{bm}
\usepackage{multirow}

\newcommand{\rev}[1]{{#1}}

\begin{document}

\title{Quantitative modeling of streamer discharge branching in air}

\author{Zhen Wang$^{1,3}$, Siebe Dijcks$^2$, Yihao Guo$^2$, Martijn van der Leegte$^2$, Anbang Sun$^3$, Ute Ebert$^{1,2}$, Sander Nijdam$^2$, Jannis Teunissen$^1$}
\address{$^1$Centrum Wiskunde \& Informatica (CWI), Amsterdam, The Netherlands\\
$^2$Eindhoven University of Technology, Eindhoven, The Netherlands\\
$^3$Xi'an Jiaotong University, Xi'an, 710049, China.}

\ead{jannis.teunissen@cwi.nl,s.nijdam@tue.nl}


\begin{abstract}
  Streamer discharges are the primary mode of electric breakdown of air in lightning and high voltage technology.
  Streamer channels branch many times, which determines the developing tree-like discharge structure.
  \rev{Understanding these branched structures is for example important to describe streamer coronas in lightning research.
  We simulate branching of positive streamers in air using a 3D fluid model where photoionization is included as a discrete and stochastic process}.
\rev{The probability and morphology of branching are in good agreement with dedicated experiments.
  This demonstrates that photoionization indeed provides the noise that triggers branching, and we show that branching is remarkably sensitive to the amount of photoionization.
  Our comparison is therefore one of the first sensitive tests for Zheleznyak's photoionization model, confirming its validity.
}
\end{abstract}

\maketitle




\ioptwocol

\section{Introduction}
\label{sec:introduction}

Streamer discharges are the first stage of electric breakdown of air (or of other gases) when suddenly exposed to high electric fields~\cite{nijdam2020physics}. They are elongated growing plasma channels; therefore their interior is largely screened from the electric field while the field is strongly enhanced at their propagating tips.
Electron impact ionization in this enhanced field causes non-linear growth with velocities of $10^5$--$10^{7} \, \mathrm{m/s}$.
Streamers are precursors of sparks and lightning leaders, they can be observed directly as sprites high above thunderclouds~\cite{cummerSubmillisecondImagingSprite2006,mcharg2007observations,luqueEmergenceSpriteStreamers2009}, and they play a prominent role in lightning inception \cite{risonObservationsNarrowBipolar2016, sterpkaSpontaneousNatureLightning2021a}. They are also widely used in plasma and high voltage technology~\cite{nijdam2020physics,fridman2005non,bruggeman2017foundations,Wang_2020}.

Branching is an integral part of streamer dynamics, as we illustrate with three examples. First, sprite discharges high above thunderstorms have been observed to start from a single channel shooting downwards from the lower edge of the ionosphere~\cite{cummerSubmillisecondImagingSprite2006, luqueEmergenceSpriteStreamers2009}; this primary streamer discharge rapidly branches out into a multi-branch tree structure over tens of kilometers. Second, similar discharge trees are seen in experiments starting from needle electrodes; they are much smaller and occur at much higher pressure, and they are related to sprites by approximate scaling laws~\cite{liuEffectsPhotoionizationPropagation2004, nijdam2020physics}. Third, radio measurements of lightning initiation in thunderstorms are interpreted as ``a volumetric system of streamers'' growing over lengths of tens to hundred of meters \cite{risonObservationsNarrowBipolar2016}. Such dynamics has recently been observed in greater detail~\cite{sterpkaSpontaneousNatureLightning2021a}, where the radio emission of the initiating discharge grew exponentially in time while the velocity was fairly constant. As sketched in the outlook of \cite{Xiaoran-steady}, the explanation could be a dynamics where streamers accelerate and become wider, and branch whenever they reach a critical radius. As streamer velocity is related to radius, the streamers would then increase exponentially in number due to repetitive branching, but move with the same average velocity.

To understand these observations and to predict multi-streamer behavior by macroscopic breakdown models~\cite{niemeyerFractalDimensionDielectric1984, luqueGrowingDischargeTrees2014}, streamer branching needs to be characterized quantitatively. Experimental methods to measure streamer branching have been developed in~\cite{Briels_2008a,Nijdam_2008,Heijmans_2013,Chen_2018}.
\rev{Here we present} fully three-dimensional simulations based on \rev{tabulated} microscopic \rev{parameters and compare them with dedicated experiments under the same conditions.} 
Our focus is on positive streamers as they emerge and propagate more easily than negative ones. They carry a positive head charge and propagate against the electron drift direction.
\\

\begin{figure}
  \centering
  \includegraphics[width=\columnwidth]{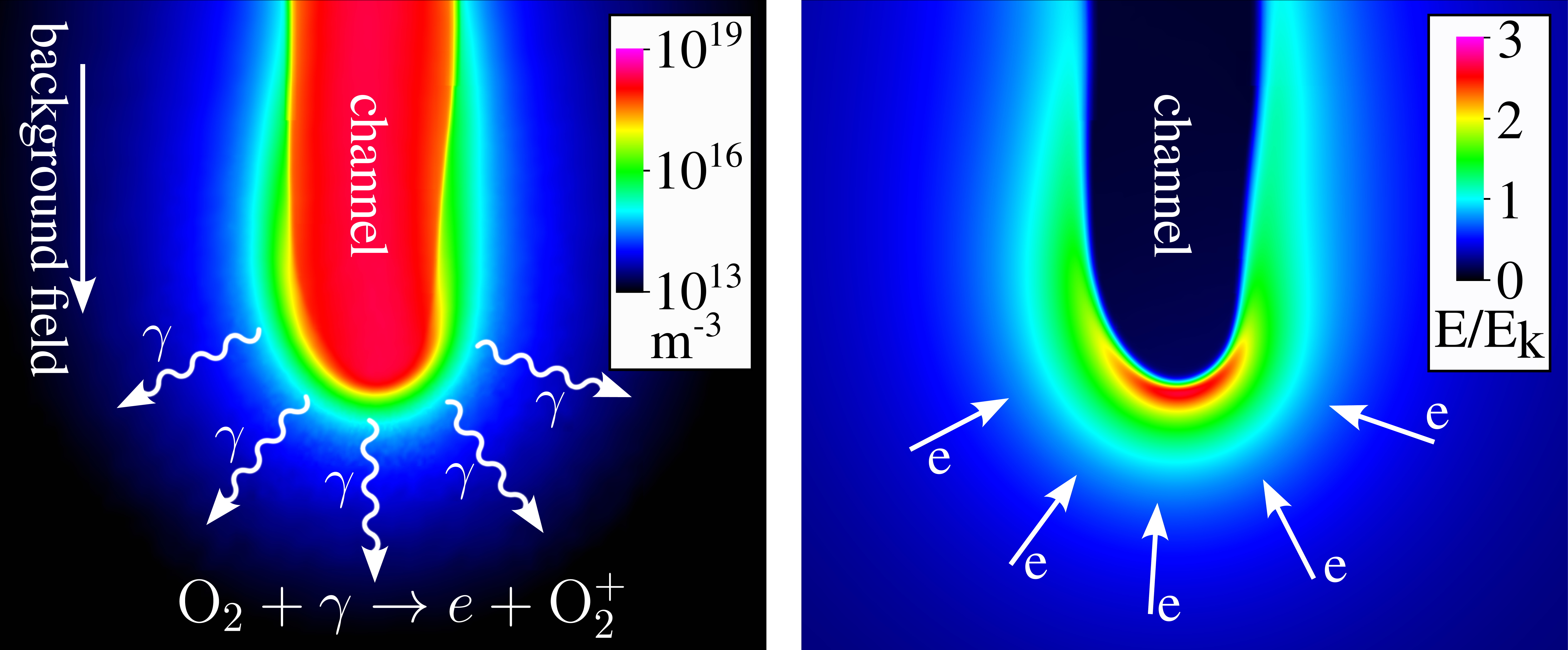}
  \caption{Cross sections through a positive streamer simulation at 15\,kV. Left: electron density, with UV photons ($\gamma$) schematically illustrated.  Right: electric field strength, relative to breakdown field $E_k$. The drift of free electrons produced by photoionization is illustrated by arrows. These electrons trigger overlapping electron avalanches propagating towards the streamer head.}
  \label{fig:streamer-growth}
\end{figure}



\section{Photoionization and branching}
\label{sec:phot-branch}

Positive streamers require seed electrons ahead of them, which in air are typically provided by photoionization~\cite{Pancheshnyi_2014,Stephens_2016}: an excited nitrogen molecule emits a UV photon that ionizes an oxygen molecule at some distance.
The liberated electrons generate electron avalanches in the high-field region in front of a streamer, which cause the streamer to grow, as illustrated in Fig.~\ref{fig:streamer-growth}.
The electron density ahead of the discharge affects the number of overlapping avalanches and thus the stochasticity of the streamer's growth.
It has been experimentally confirmed that there is more branching in gases with less photoionization and less background ionization, see e.g.,~\cite{Nijdam_2010,Nijdam_2011b,Takahashi_2011}.

That the stochasticity of photoionization triggers branching is also found in simulations in 2D~\cite{Xiong_2014}, and in full 3D~\cite{teunissen20163d, bagheri2019effect, Marskar_2020}, while early 3D studies~\cite{Pancheshnyi_2005a} worked with stochastic background ionization.
  Branching simulated in~\cite{teunissen20163d, bagheri2019effect, Marskar_2020} qualitatively resembled branching in experiments, but no quantitative comparison was performed -- \rev{
this is the goal of the present paper.}

In general, protrusions in the space charge layer around a streamer head can locally enhance the electric field, causing them to grow.
This Laplacian destabilization can occur in a fully deterministic manner~\cite{Arrayas2002, Ebert_2010}, but it is accelerated by noise~\cite{luqueElectronDensityFluctuations2011}.
\\

\section{Set-up of experiments and simulations}
\label{sec:set-up-experiments}

\begin{figure}
  \centering
  \includegraphics[width=\columnwidth]{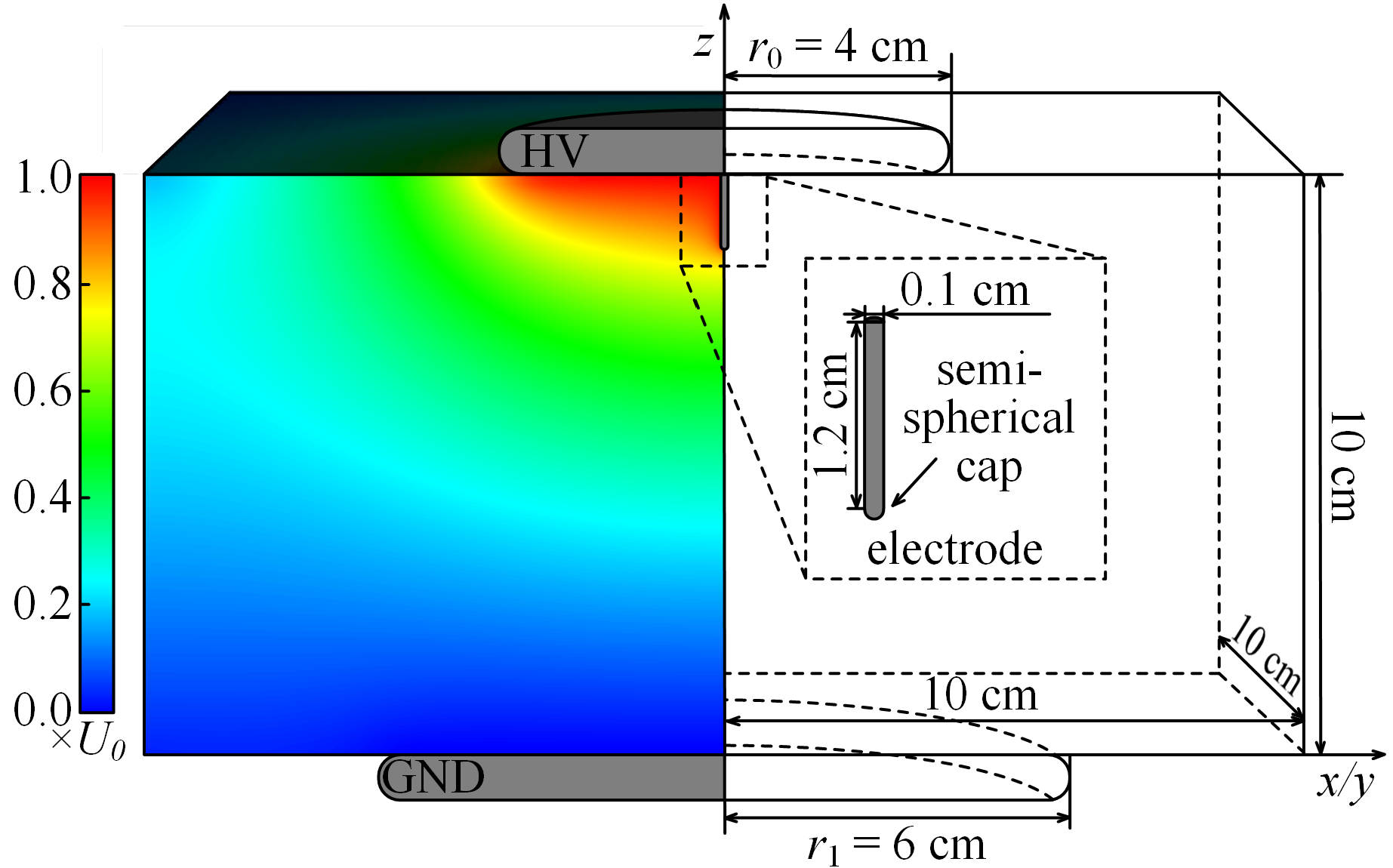}
  \caption{Electrode geometry both in simulations and experiments.
    The full computational domain is $20 \, \mathrm{cm}\times 20 \, \mathrm{cm}\times 10 \, \mathrm{cm}$; half of it is shown.
    There are plate electrodes at the upper and lower boundaries. The discharges start from a needle electrode that protrudes from the upper electrode.
    The electric potential distribution without space charge is shown on the left.
    In the experiments, the electrodes are inside a grounded discharge vessel.
    In the simulations, custom boundary conditions for the electric potential are used to account for this vessel, as described in~\cite{Li_2021}.}
  \label{fig:intial-condition}
\end{figure}

To obtain a more quantitative understanding, we here compare streamer branching in simulations and experiments under the same conditions.
The simulations and experiments are performed in synthetic air (80\%\,$\mathrm{N_2}$, 20\%\,$\mathrm{O_2}$, no humidity) at 233\,mbar and approximately 300\,K, under applied voltages of 15\,kV, 17\,kV and 19\,kV, using the geometry illustrated in Fig.~\ref{fig:intial-condition}. Under these conditions, experiments with a moderate amount of branching could be performed, which could also be imaged well.


\begin{figure}
  \centering
  \includegraphics[width=\linewidth]{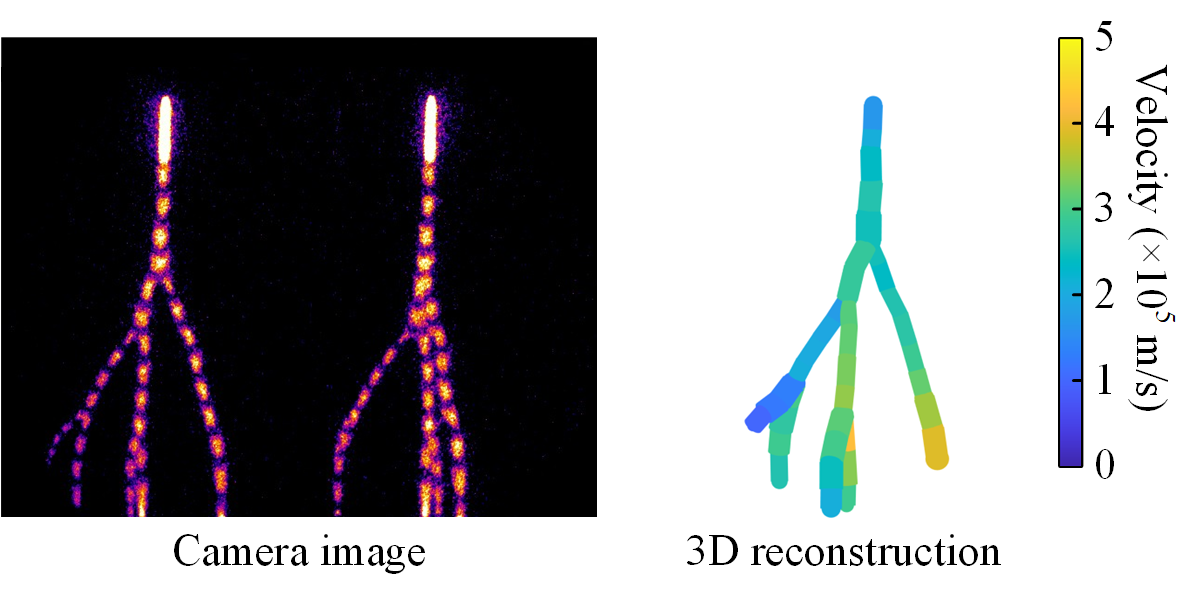}
  \caption{Example of 3D reconstruction of streamer paths and velocities in experiments, using stereoscopic stroboscopic images.}
  \label{fig:path-identification}
\end{figure}

\begin{figure*}
  \centering
  \includegraphics[width=\linewidth]{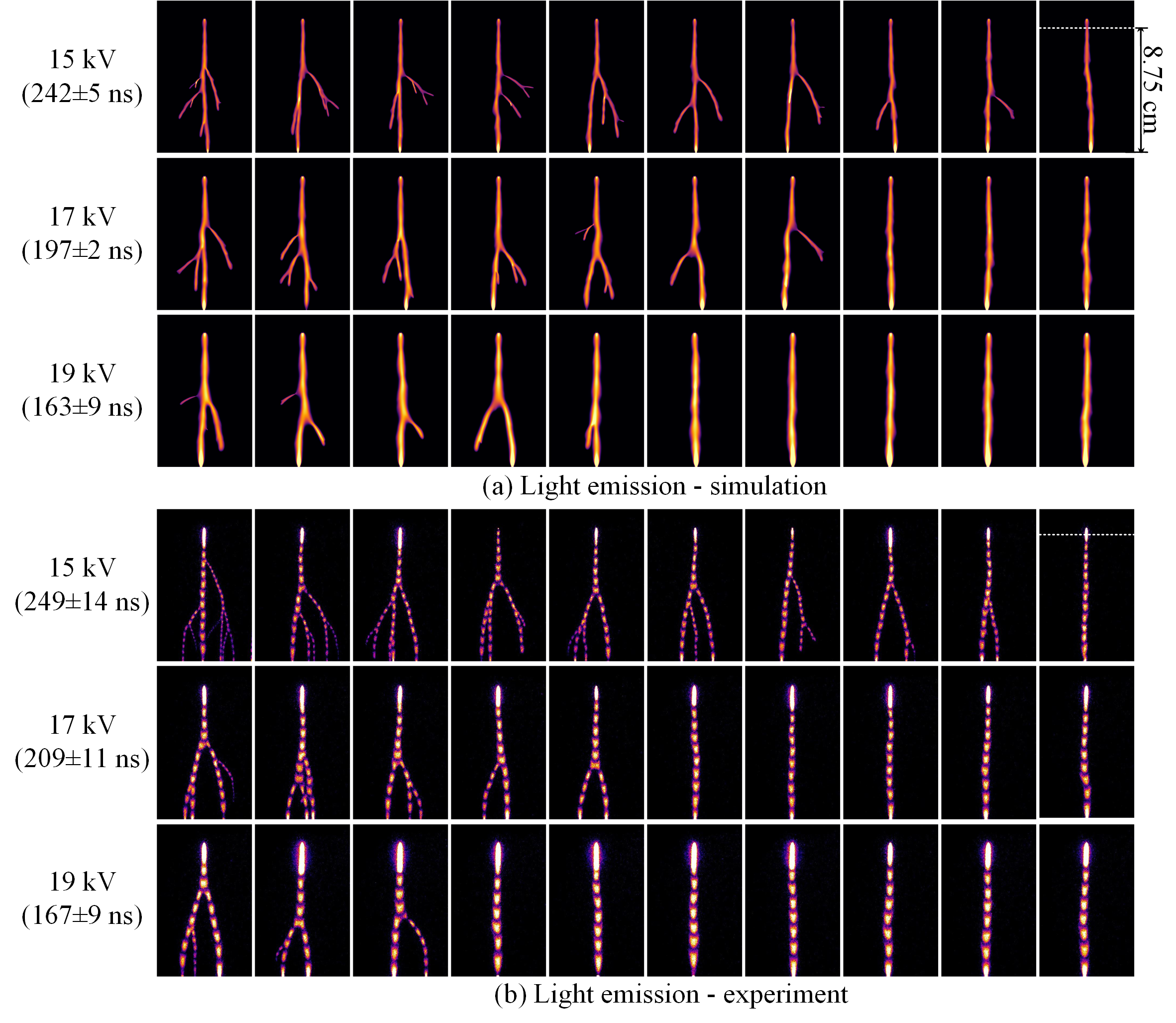}
  \caption{Comparison of streamer branching morphologies under applied voltages of 15, 17 and 19\,kV, all at 233\,mbar. For each voltage 60 simulations and 128 experiments were performed, and the 10 figures shown for each case are representative for the distribution given in table \ref{tab:probability-comparison}, with branched cases on the left.
    The simulations were stopped when the primary streamer reached the bottom electrode.
    In the experiments, a bright area is visible near the upper needle electrode due to a secondary streamer. Average times for crossing the last $8.75 \, \mathrm{cm}$ of the gap are indicated on the left, together with standard deviations.
  }
  \label{fig:morpho-comparison}
\end{figure*}

The experiments are performed with a pulse repetition rate of 20\,Hz.
Images are captured that are both stereoscopic and stroboscopic, as illustrated in Fig.~\ref{fig:path-identification}.
We use a similar stereoscopic setup as in~\cite{Nijdam_2008}.
In stroboscopic mode, the ICCD camera (LaVision PicoStar HR) has a gating time of 8\,ns and and a repetition rate of 50\,MHz.
From the captured images, 3D paths of streamers are reconstructed.
This is done by connecting the bright dots, resulting from the stroboscopic gating, based on a shortest-path tree algorithm that can account for streamer branching.
A quadratic extrapolation is used to smooth the streamer paths, from which branching angles and local velocities are obtained.
More detailed information about this scheme can be found in~\cite{Dijcks_2023}.

\begin{table}
	\centering
	\caption{Reactions included in the model. Rate coefficients for $k_1$ to $k_5$ were computed using BOLSIG+~\cite{hagelaar2005solving,bolsig2019} from Phelps' cross sections~\cite{phelps1985anisotropic,Phelps-database}, and $k_6$ to $k_8$ were obtained from~\cite{Pancheshnyi_2005}.}
	\begin{tabular}{ll}
          \hline
          \hline
		Reaction & Rate coefficient \\ \hline
		$\textrm{e} + \textrm{N}_2 \stackrel{k_1}{\longrightarrow} \textrm{e} + \textrm{e} + \textrm{N}_2^+$ & $k_1(E/N)$ \\
		$\textrm{e} + \textrm{O}_2 \stackrel{k_2}{\longrightarrow} \textrm{e} + \textrm{e} + \textrm{O}_2^+$ & $k_2(E/N)$ \\
		$\textrm{e} + \textrm{O}_2 + \textrm{O}_2 \stackrel{k_3}{\longrightarrow} \textrm{O}_2^- + \textrm{O}_2$ 	& $k_3(E/N)$  \\
		$\textrm{e} + \textrm{O}_2 \stackrel{k_4}{\longrightarrow} \textrm{O}^- + \textrm{O}$ 			& $k_4(E/N)$  \\
		$\textrm{e} + \textrm{N}_2 \stackrel{k_5}{\longrightarrow} \textrm{e} + \textrm{N}_2(\textrm{C}^3 \Pi_u)$ 	& $k_5(E/N)$ \\
		$\textrm{N}_2(\textrm{C}^3 \Pi_u) + \textrm{N}_2 \stackrel{k_6}{\longrightarrow} \textrm{N}_2 + \textrm{N}_2$ & $k_6 = 0.13\times10^{-16}\,\textrm{m}^{3}\textrm{s}^{-1}$  \\
		$\textrm{N}_2(\textrm{C}^3 \Pi_u) + \textrm{O}_2 \stackrel{k_7}{\longrightarrow} \textrm{N}_2 + \textrm{O}_2$ & $k_7 = 3.0\times10^{-16}\,\textrm{m}^3\textrm{s}^{-1}$  \\
		$\textrm{N}_2(\textrm{C}^3 \Pi_u)\stackrel{k_8}{\longrightarrow} \textrm{N}_2(\textrm{B}^3 \Pi_g)$ 		& $k_8=1/(42\,\textrm{ns})$  \\
          \hline
          \hline
	\end{tabular}
	\label{tbl:reaction_table}
\end{table}

Simulations are performed with a 3D drift-diffusion-reaction fluid model in which the only source of stochasticity is \rev{the discreteness of} photoionization.
We have recently established the approximate validity of this model for propagating streamers by comparing against experimental results~\cite{Li_2021} and particle simulations~\cite{Wang_2022}.
The model is described in detail in~\cite{teunissen2017simulating,bagheri2019effect,Li_2021,Wang_2022}, but we provide a brief overview below.
The electron de nsity $n_e$ evolves in time as
\begin{equation}
  \partial_t n_e =  \nabla \cdot (n_e\mu _e\mathbf{E} + D_e\nabla n_e) + S_i-S_a + S_\mathrm{ph},
  \label{eq:fluid-model_ne}
\end{equation}
where $\mu_e$ and $D_e$ are the electron mobility and the diffusion coefficient, $S_\mathrm{ph}$ is the non-local photoionization source term discussed below, \rev{and $S_i - S_a$ is a source term due to the ionization ($S_i$) and attachment ($S_a$) reactions given in table~\ref{tbl:reaction_table}.}
Electron transport coefficients are assumed to be functions of the local electric field.
They are computed from electron-neutral cross sections for $\mathrm{N_2}$ and $\mathrm{O_2}$~\cite{phelps1985anisotropic,Phelps-database} using BOLSIG+~\cite{hagelaar2005solving,bolsig2019}.
\rev{Ions and neutral species are assumed to be immobile, and their densities $n_j$ (for $j = 1, 2, \dots$) evolve as
\begin{equation}
  \partial_t n_j =  S_j,
  \label{eq:fluid-model_ni}
\end{equation}
with $S_j$ determined by the reactions from table~\ref{tbl:reaction_table}.}

At every time step, the electric field is computed as ${\bf E} = -\nabla \phi$,
where the electric potential $\phi$ is obtained by solving Poisson's equation~\cite{teunissen2017simulating,Teunissen_2022}.
For $\mathrm{N}_2$-$\mathrm{O}_2$ mixtures close to atmospheric pressure, the $\mathrm{N}_2(\mathrm{C}^3\Pi_u \to \mathrm{B}^3\Pi_g$) transition is the main source of emitted light~\cite{pancheshnyi2000discharge}.
In the simulations, we approximate the time-integrated light emission by the time integral over this transition.

For photoionization, a Monte-Carlo version of Zheleznyak's model~\cite{Zheleznyak_1982} with discrete photons is used, as described in~\cite{chanrion2008pic,bagheri2019effect}.
\rev{The photo-ionization source
term $S_{\mathrm{ph}}(r)$ is then given by
\begin{equation}
  {S_\mathrm{ph}}(\mathbf{r}) =
  \int{\frac{I(\mathbf{r'})f(\left|{\mathbf{r}-\mathbf{r'}}\right|)}{4\pi\left|{\mathbf{r}-\mathbf{r'}}\right|^2} d^3\mathbf{r'}},
     \label{eq:photoionization}
\end{equation}
where $f(r)$ is the photon absorption function~\cite{Zheleznyak_1982} and $I(\mathbf{r})$ is the source of ionizing photons, which is proportional to the electron impact ionization source term $S_i$:
\begin{equation}
  I(\mathbf{r}) = \frac{p_q}{p+p_q}\xi S_i.
  \label{eq:UV-source-term}
\end{equation}
}
Here $p$ is the gas pressure, $p_q = 40 \, \mathrm{mbar}$ is the quenching pressure and $\xi$ a proportionality factor.
In principle, $\xi$ depends on the electric field~\cite{Zheleznyak_1982}, but we here for simplicity approximate it by a constant $\xi = 0.075$~\cite{bagheri2019effect}.
In each computational grid cell, \rev{the number of emitted photons is sampled from a Poisson distribution with the mean given by $I(\mathbf{r}) \Delta t \Delta V$, where $\Delta t$ is the time step and $\Delta V$ is the volume of the cell.}
For each ionizing photon, an isotropic angle and an absorption distance (according to Zheleznyak {\it et al.}~\cite{Zheleznyak_1982}) are sampled.
The photons are then absorbed on the numerical grid to determine the photoionization source term $S_\mathrm{ph}$.


In the experiments, the voltage rise time was about 100 ns, but inception would typically occur with a delay of several hundred ns, when the voltage had already reached its maximum.
  To ensure a significant probability of inception, a voltage pulse width and a camera gate time of $1 \, \mu\mathrm{s}$ were used.
  In the simulations, we therefore do not take the voltage rise time into account, but instead apply a constant voltage from time zero.
  A homogeneous background ionization density of $10^{11}\, \mathrm{m^{-3}}$ of electrons and positive ions is included to facilitate discharge inception.
  This density has no significant effect on the later discharge propagation since photoionization produces ionization densities that are orders of magnitude higher\rev{~\cite{Wormeester_2010}, as also illustrated in~\ref{sec:ionization-density-estimate}.}

\section{Results}
\label{sec:results}

For each applied voltage, 60 3D simulations were performed and 128 experimental images were captured.
Figures~\ref{fig:morpho-comparison}(a) and \ref{fig:morpho-comparison}(b) show ten representative examples from simulations and experiments for each voltage.
The number of (non-)branching cases shown is proportional to the measured branching percentages as given in Table~\ref{tab:probability-comparison}.

\begin{table}
    \centering
    \caption{The number of cases with and without branching versus applied voltage. \rev{For the branching percentages, an estimate of the standard deviation due to the limited sample size is included. Cases without inception are excluded from the branching statistics.}
  }
    \begin{tabular}{C{0.5cm}C{2.2cm}C{1.2cm}C{1.2cm}C{1.2cm}}
     \hline
     \hline
     &                       & 15 kV & 17 kV & 19 kV               \tabularnewline
     \hline
     \multirow{3}{*}{Sim.}
     & Branched    & 55    & 46    & 30                  \tabularnewline
     & Non-branched       & 5     & 14    & 30                  \tabularnewline
     & Branched \% & $92\pm$ 4\%  & $77\pm 5$\% & $50\pm 6$\%               \tabularnewline
     \hline
     \multirow{4}{*}{Exp.}
     & Branched    & 34    & 60    & 40                  \tabularnewline
     & Non-branched       & 2     & 54    & 78                  \tabularnewline
     & No inception          & 92    & 14    & 10                  \tabularnewline
     & Branched \% & $94\pm 4$\%  & $53\pm 5$\%  & $34\pm 4$\%                \tabularnewline
     \hline
     \hline
   \end{tabular}
   
   \label{tab:probability-comparison}
\end{table}

The morphology of the simulated and experimental discharges is highly similar. \rev{The branching angles, the location of first branching, and the streamer optical radii all agree well.
The percentage of cases in which the primary streamer branches differs up to a factor of about $1.5$ between experiments and simulations, but we argue below that this is still very good agreement given the sensitivity of this percentage to the photoionization coefficients.}
The average time it takes streamers to cross the last $8.75 \, \mathrm{cm}$ of the gap is indicated in figure~\ref{fig:morpho-comparison}.
  These gap bridging times agree within about 5\% between simulations and experiments, and in both cases they were similar for branched and non-branched cases.
  Streamer velocities ranged from about $0.3 \, \mathrm{mm/ns}$ to $0.6 \, \mathrm{mm/ns}$, with average velocities in the second half of the gap being about 20-25\% higher than in the first half.

\begin{figure}
  \centering
  \includegraphics[width=\linewidth]{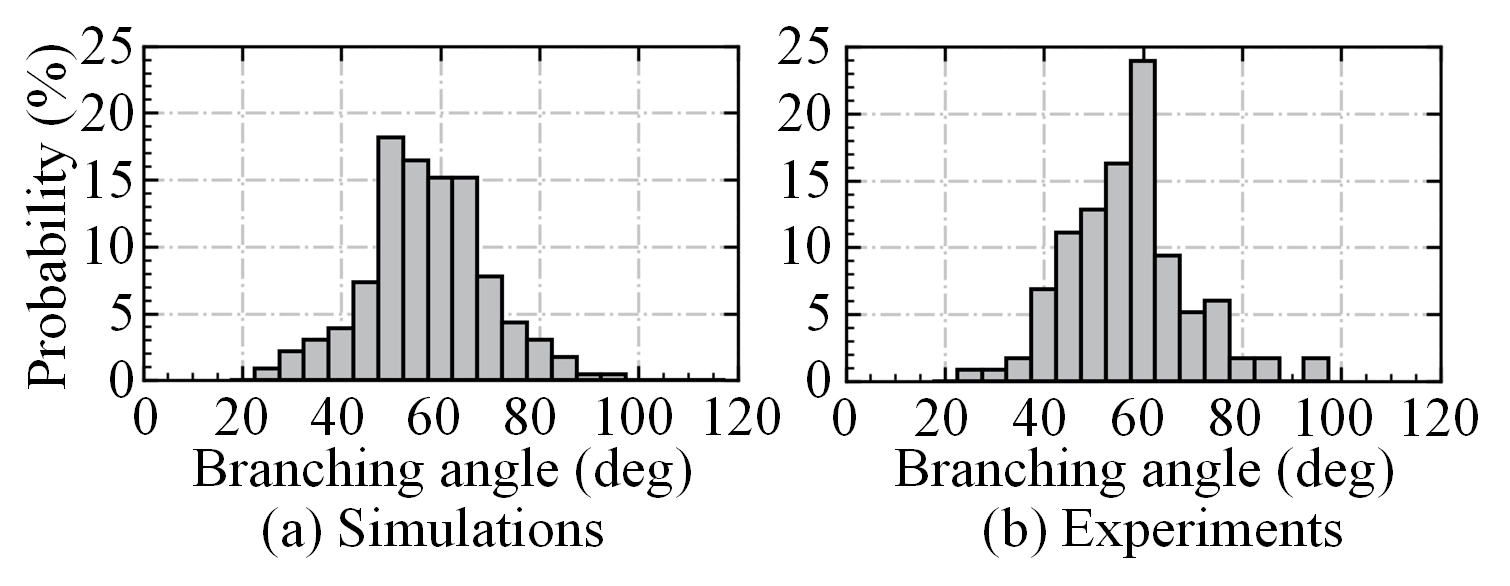}
  \caption{Probability distribution of the angle between two new segments after branching.}
  \label{fig:branching-angle}
\end{figure}

\begin{figure}
  \centering
  \includegraphics[width=\linewidth]{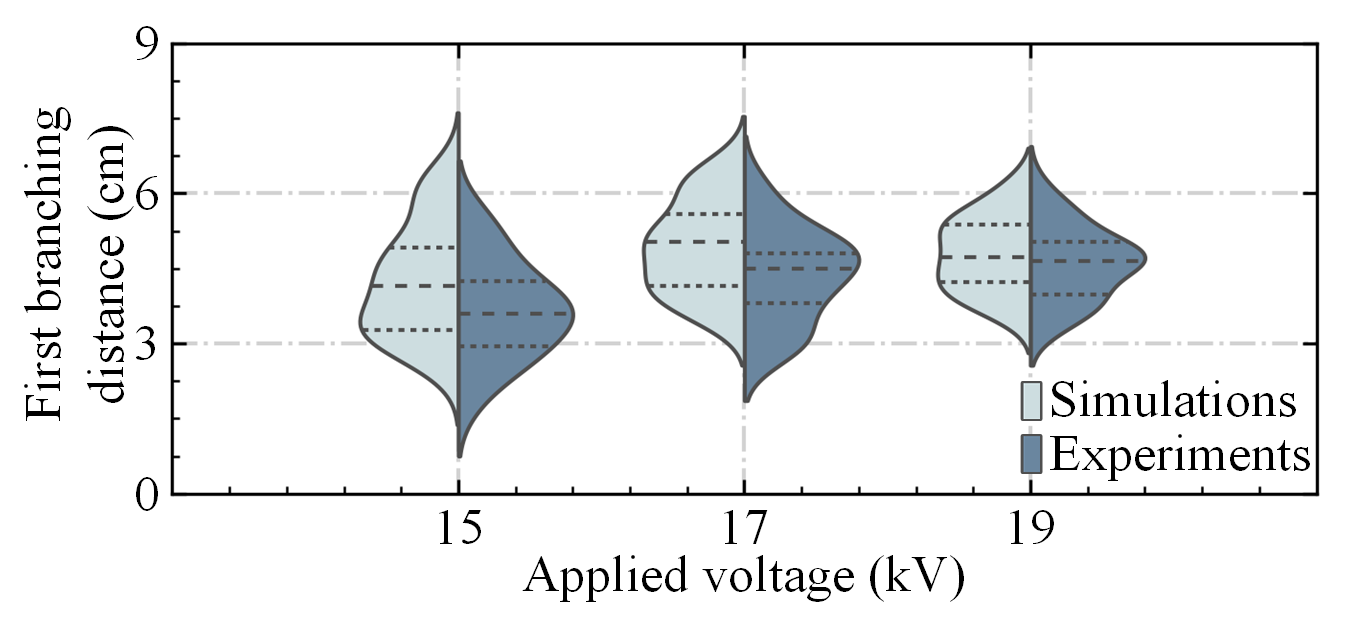}
  \caption{Distributions of the distance until a first branching, as measured from the electrode tip. The horizontal dashed lines indicate quartiles. A kernel density estimation of the underlying data is also shown. (The area between quartiles is not conserved due to smoothing.)}
  \label{fig:first-branching-point}
\end{figure}

Figure~\ref{fig:branching-angle} shows the distribution of branching angles, measured between the two new segments.
The mean branching angle was $60^\circ$ in the simulations and $58^\circ$ in the experiments, with respective standard deviations of $16.1^\circ$ and $12.0^\circ$.
The distribution of the first branching location is shown in Fig.~\ref{fig:first-branching-point}.

\begin{table}
    \centering
    \caption{The sensitivity of streamer branching to the photoionization coefficient $\xi$ in equation~(\ref{eq:UV-source-term}).
      The simulations were performed at 17\,kV.
      $N_\mathrm{branchings}$ denotes the average number of branching events.
      Experimental values are included for comparison.}
    \begin{tabular}{C{2.0cm}C{1.0cm}C{1.0cm}C{1.0cm}|C{1.0cm}}
     \hline
     \hline
     $\xi$ & 0.0375 & 0.075 & 0.15 & Exp. \tabularnewline
     \hline
     Branched \%        & 85\%   & 77\%  & 5\% & 53\%       \tabularnewline
     $N_\mathrm{branchings}$    & 7.30    & 1.40   & 0.05 & 0.92       \tabularnewline
     \hline
     \hline
    \end{tabular}
    \label{tab:xi-sensitivity}
\end{table}

\begin{figure}
  \centering
  \includegraphics[width=\linewidth]{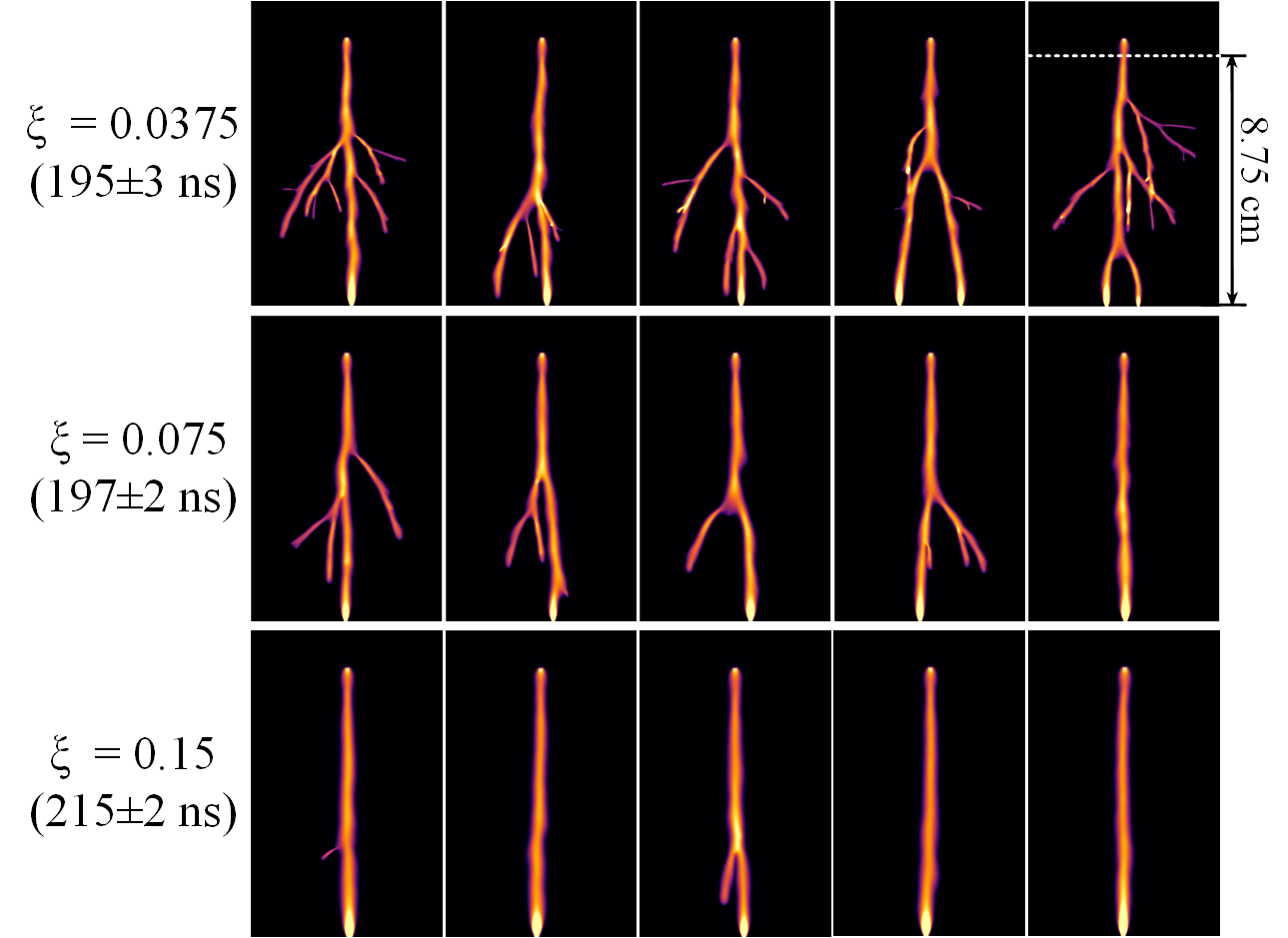}
  \caption{Representative simulations of streamer branching for different photoionization coefficients $\xi$. The value of $\xi$ for each row is given on the left, and the value used elsewhere in the paper is $\xi = 0.075$. The simulations were performed at 17\,kV.
    Experimental images at 17\,kV are shown in figure \ref{fig:morpho-comparison}.}
  \label{fig:xi-sensitivity}
\end{figure}

\rev{
  As the applied voltage increases, the percentage of cases in which the primary streamer branches decreases.
  The reason for this is that more ionization is produced at a higher voltage, and thus also more photoionization, which makes the growth of the streamer less stochastic.}
At 15 kV, the branching percentage is almost the same in experiments and simulations.
At 17~kV and 19~kV, the branching percentage is about $1.5$ times larger in the simulations.
We consider this good quantitative agreement, since the branching probability in simulations is very sensitive to the photoionization coefficients.
To demonstrate this sensitivity, we have varied the parameter $\xi$ in equation~(\ref{eq:UV-source-term}), by setting it to half and double the value of $\xi = 0.075$ used elsewhere in the paper.
The resulting branching statistics are described in Table~\ref{tab:xi-sensitivity}, and representative cases are shown in Fig.~\ref{fig:xi-sensitivity}.
When halving or doubling $\xi$, the branching behavior qualitatively and quantitatively disagrees with the experiments.
In contrast, average streamer velocities (deduced from the gap bridging times in Fig.~\ref{fig:xi-sensitivity}) are not sensitive to $\xi$.
  When $\xi$ is halved, there is hardly any difference, and when $\xi$ is doubled the velocity is about 10\% lower.

Zheleznyak's photoionization model is a rather simple approximation of several photoionization mechanisms~\cite{Stephens_2016}, in which the coefficient $\xi$ is essentially a fitting parameter.
  In~\cite{Pancheshnyi_2014}, it was pointed out that $\xi$ can vary between about 0.02 and 0.2 in air, depending on the electric field strength and the experimental data used for the fit.
  Given these uncertainties, and given the sensitivity of the simulations with respect to $\xi$, we think the agreement between simulations and experiments is surprisingly good.
  We furthermore emphasize that the constant value $\xi = 0.075$ used here was based on previous work~\cite{bagheri2019effect} and not tuned in any way.
  Our results therefore suggest that Zheleznyak's model gives an accurate description of photoionization in air.
  \\




\section{Conclusions}
\label{sec:conclusions}

We have found quantitative agreement between simulations and experiments of positive streamer branching in air, from which we draw three main conclusions: First, we have demonstrated that photoionization is the main mechanism that governed the branching observed here, as this was the only source of stochastic fluctuations in the simulations.
Second, our comparison is one of the first sensitive tests for Zheleznyak's photoionization model, since the branching probability was shown to be very sensitive to the photoionization coefficients, \rev{whereas other streamer properties like velocity are much less sensitive to these coefficients}.
Third, the presented validation of the model opens the opportunity to computationally study streamer branching.
This is important for understanding the physical questions addressed in the introduction, in which branching plays a fundamental role in the discharge evolution.

\vspace{1em}
\textit{Acknowledgements.} ZW was supported by internal means of CWI Amsterdam, and SD by the Netherlands' STW-project 15052 `Let CO$_2$ spark!'.
YG was supported by the China Scholarship Council (CSC) Grant No.\ 202006280041.

  \appendix

  \cleardoublepage





\section{Photoionization and initial electron density}
\label{sec:electron-density}

\begin{figure}
  \centering
  \includegraphics[width=\columnwidth]{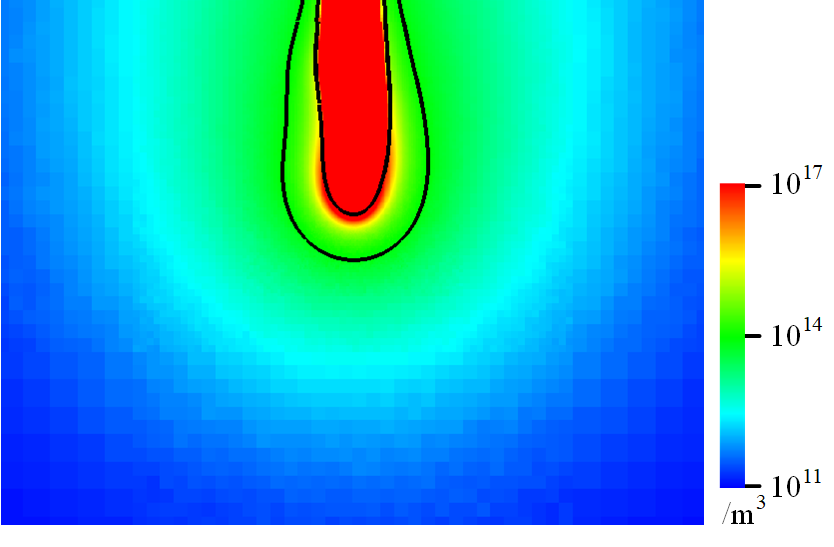}
  \caption{Cross section through a simulation at 15 kV, showing the electron density around a non-branched streamer in the middle of the discharge gap.
    The black contour lines demarcate the area in which the electric field is above breakdown.
    }
  \label{fig:stochacity}
\end{figure}
The electron density around a streamer head is illustrated in figure~\ref{fig:stochacity}, which shows a cross section through a simulation at 15\,kV.
Although photoionization was here found to be the main mechanism behind streamer branching, it can be seen that it produces a relatively smooth electron density around the streamer head.
The region where the electric field is above breakdown is indicated in the figure.
The electron density at the outer boundary of this region is about $10^{14} \, \mathrm{m}^{-3}$.
It is therefore not possible to identify particular photoionization events (or the resulting avalanches) with branching events.
Instead, fluctuations in the electron density ahead of the discharge deform the streamer head shape, and these deformations can lead to branching.
They also cause the non-straight growth of non-branched streamer channels, see for example figure~\ref{fig:time-evolution}.

\begin{figure*}
  \centering
  \includegraphics[width=\textwidth]{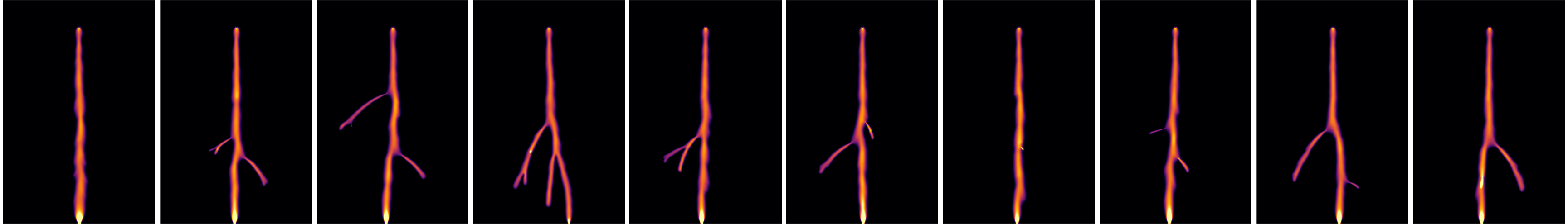}
  \caption{10 runs of streamers initiated from a Gaussian seed at 15~kV. Here the initial electron and ion densities are given by a Gaussian distribution $n_i(\mathbf{r})=n_e(\mathbf{r})= 10^{13}\,\mathrm{m}^{-3}
    \exp\left[{-(\mathbf{r}-\mathbf{r_0})^2}/(2 \, \mathrm{mm})^2)\right]$, where $r_0$ is the location of the tip of the electrode.}
  \label{fig:background_density}
\end{figure*}

Note that the electron density produced by photoionization is several orders of magnitude higher than the background electron density of $10^{11} \, \mathrm{m}^{-3}$ that was used in the simulations as an initial condition.
This background ionization therefore has no significant effect on our simulation results. This is illustrated in figure~\ref{fig:background_density}, in which it is replaced by a localized Gaussian seed \rev{with a peak density of $10^{13}\,\mathrm{m}^{-3}$.
This seed provides the first electrons near the electrode to ensure a discharge can start, but it has no significant effect on the later discharge evolution.}

\section{Ionization density due to previous pulses}
\label{sec:ionization-density-estimate}

The experiments use voltage pulses of 1~$\mu$s duration at a repetition rate of 20 Hz, so there are 50\,ms between the pulses. During this time electrons attach to oxygen, forming negative ions, and positive and negative ions recombine.
If effects due ion diffusion are ignored, the ion density $n$ at the start of a next pulse can be estimated as~\cite{Pancheshnyi_2005a,Wormeester_2010,Kossyi_1992}:
\begin{equation}
  \label{eq:pancheshnyi}
  n(t) = (k_\mathrm{rec} \, t)^{-1},
\end{equation}
where $k_\mathrm{rec}$ is the effective ion recombination rate, which typically lies between $10^{-12} \, \mathrm{m}^3\mathrm{s}^{-1}$ and $10^{-13} \, \mathrm{m}^3\mathrm{s}^{-1}$~\cite{Kossyi_1992}.
This gives an estimated ionization density $n(50\,\mathrm{ms})$ between $2\times 10^{13} \, \mathrm{m}^{-3}$ and $2\times 10^{14} \, \mathrm{m}^{-3}$.
These densities are comparable to the electron density produced by photoionization, see figure~\ref{fig:stochacity}.
If the main negative ions would for example be $\mathrm{O}_2^-$ or $\mathrm{O}^-$, then they could have a significant effect on the next pulse due to electron detachment.

However, previous work on discharge inception~\cite{Mirpour_2020} has indicated that remaining negative ions do not easily give up electrons through detachment.
This is consistent with the fact that inception often occurred with a significant delay in our experiments.
A possible explanation could be that the main stable negative ion is $\mathrm{O}_3^-$~\cite{Kossyi_1992,Pancheshnyi_2013}, from which electrons hardly detach.
We therefore expect background ionization from previous pulses to not significantly affect the branching behavior observed here. For more recent results on the effect of ion conversion and 
of electron attachment and detachment processes on the electron density in 
repetitive discharges, we refer to~\cite{francisco2021electrically,guo2023computational,malla2023double}.

\section{\rev{Pulse rise time}}
\label{sec:rise_time}\

    \begin{figure*}
        \centering
        \includegraphics[width=\textwidth]{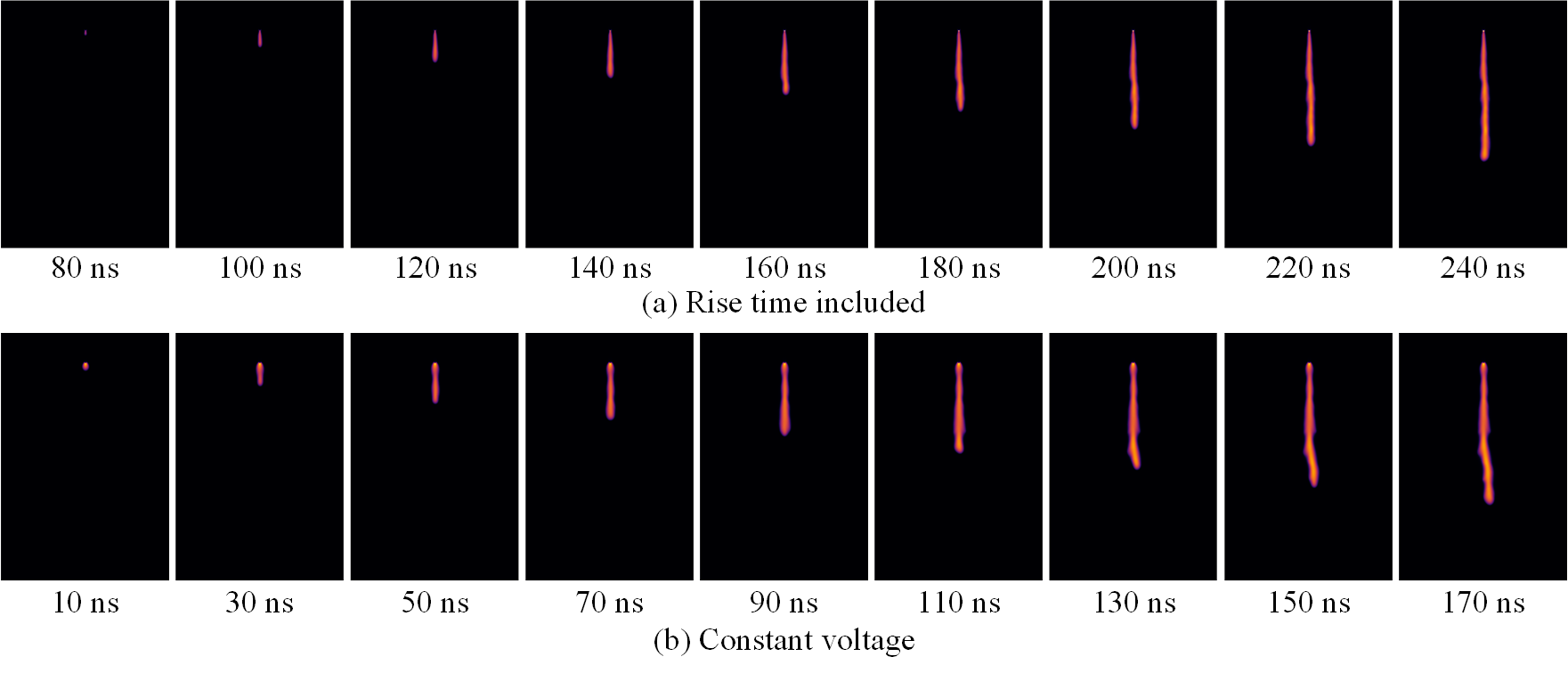}
        \caption{Time evolution in simulations at 15~kV.
          For the top row a rise time of 105 ns was used, for the bottom row the voltage was applied instantaneously.
        }
        \label{fig:rising_time}
\end{figure*}

\rev{In the experiments, a rise rate of $0.14 \, \textrm{kV/ns}$ was used for the different applied voltages, which leads to rise times of about 105~ns (at 15~kV), 119~ns (at 17~kV) and 133~ns (at 19~kV).
  As discussed in the main text, inception typically occurred when the voltage had already reached its maximum, which is why in the simulations the rise time was not taken into account.
  We now briefly test how the inclusion of a finite rise time affects the simulation results.

  Figure~\ref{fig:rising_time} shows examples of streamer evolution with and without a rise time at 15~kV. Note that streamer inception occurs around 100~ns, when the applied voltage is already about 15~kV, so that the main effect is simply a delay in streamer inception.
  We observed similar inception delays of about 100~ns at voltages of 17~kV and 19~kV.
  The reason the rise time has no significant effect on the later propagation is that these voltages are all rather close to the inception voltage.
If we would apply a significantly higher voltage the streamer would already propagate a significant distance while the voltage was rising, leading to a stronger dependence on the rise time~\cite{Komuro_2021}.
  }

\section{Time evolution}
\label{sec:evolution}

\begin{figure*}
  \centering
  \includegraphics[width=\textwidth]{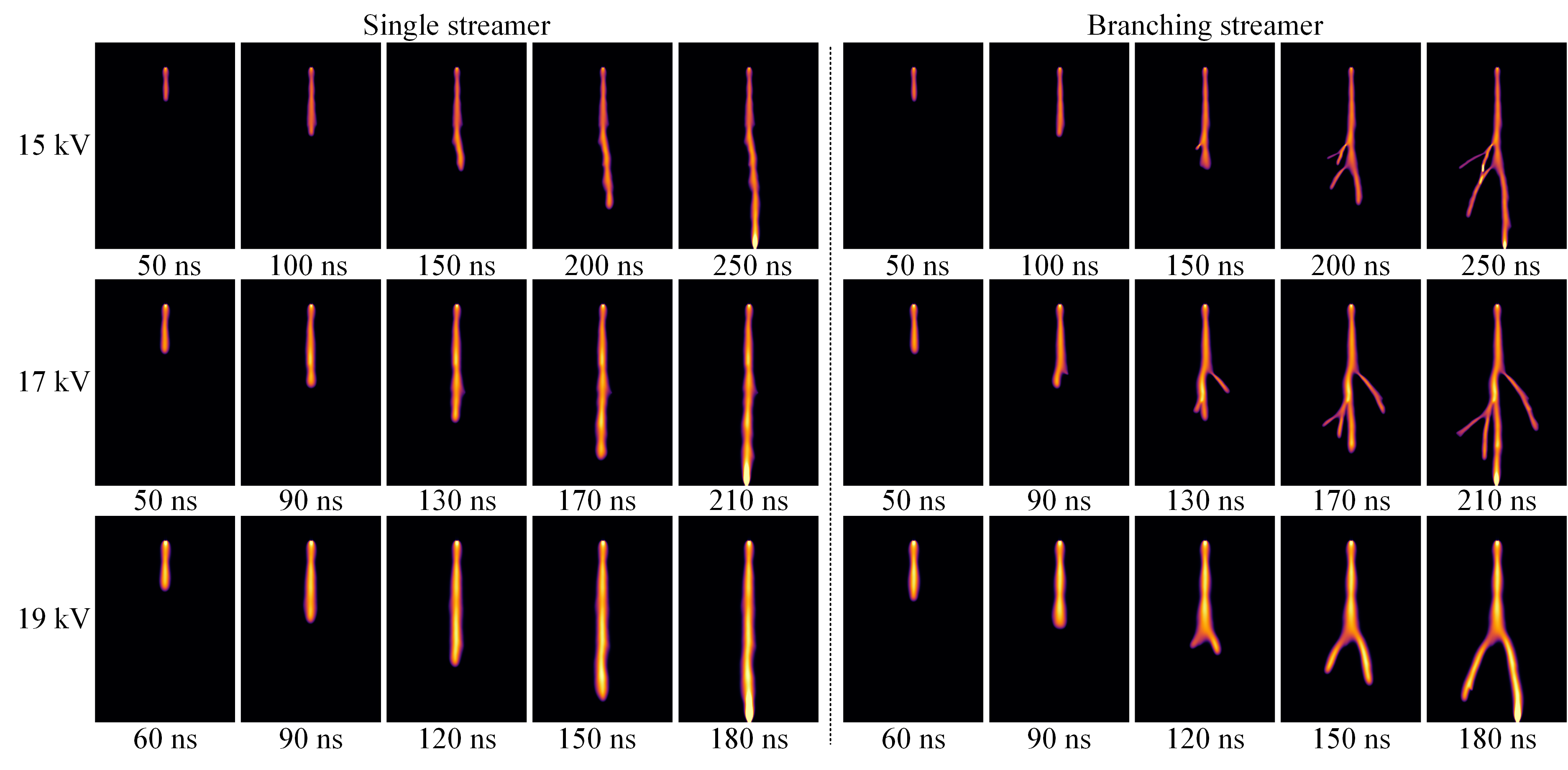}
  \caption{Examples of time evolution in simulations under applied voltage of 15 kV, 17 kV and 19~kV.
    Shown is the integrated light emission, with cases without branching on the left and cases with branching on the right.
  }
  \label{fig:time-evolution}
\end{figure*}

Figure~\ref{fig:time-evolution} illustrates the time evolution in simulations at different applied voltages.
At each voltage, both single and branching streamers bridge the gap around the same time, so branching does not significantly affect the streamer velocity, as also discussed in the main text.

\section{Radii before and after branching}
\label{sec:radii}

\begin{figure}
  \centering
  \includegraphics[width=\columnwidth]{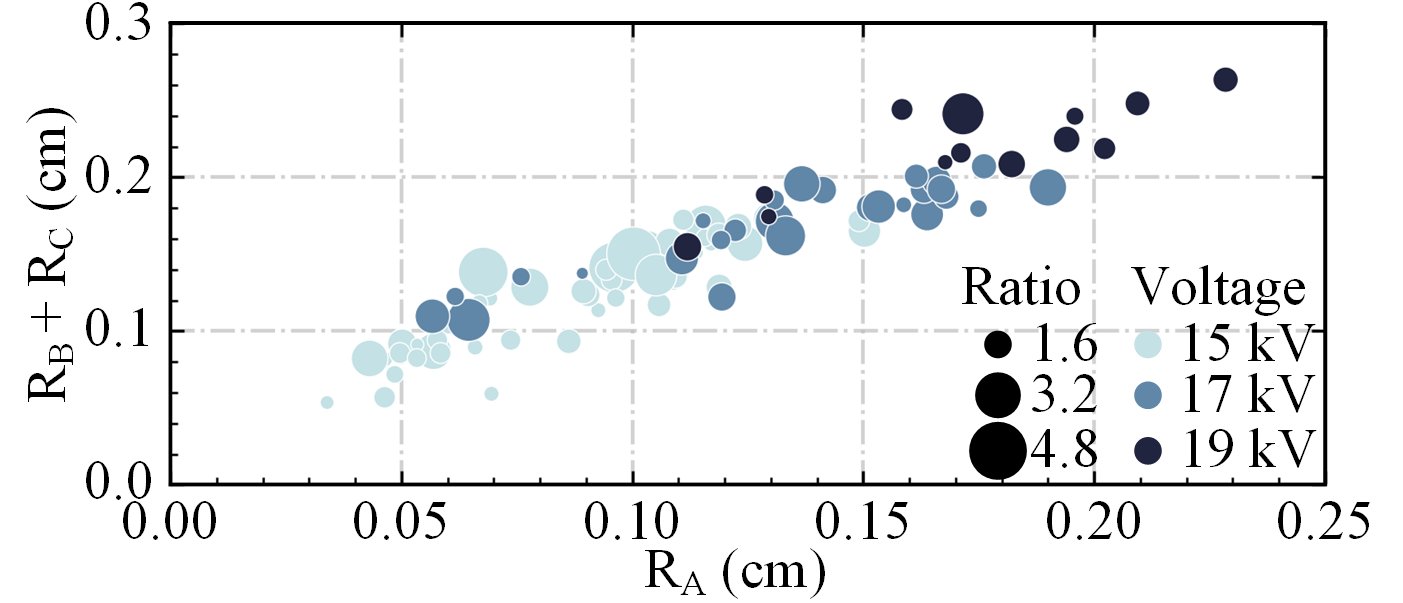}
  \caption{Streamer radii 20\,ns before ($\mathrm{R_A}$) and 20\,ns after ($\mathrm{R_B}$, $\mathrm{R_C}$) branching in simulations.
    The sizes of the circles represent the ratio $\mathrm{R_B}/\mathrm{R_C}$, with $\mathrm{R_B} \geq \mathrm{R_C}$.}
  \label{fig:cs-conservation}
\end{figure}

In the simulations, we have measured streamer radii before and after branching.
Figure~\ref{fig:cs-conservation} shows the sum of the radii after branching ($\mathrm{R_B} + \mathrm{R_C}$) versus the parent radius $\mathrm{R_A}$.
The results suggest a relation $\mathrm{R_A}=\mathrm{k}\times(\mathrm{R_B} + \mathrm{R_C})$, with $\mathrm{k} \approx 1.3$, but they are also consistent with the relation $\mathrm{R_A^2}=\mathrm{R_B^2}+\mathrm{R_C^2}$ observed before in~\cite{Nijdam_2008,Chen_2018}.

\section{Computional cost}

Typical computing times for a single run under the conditions of the main text were 12 to 36 hours. These computations ran on Snellius, the Dutch national supercomputer, using 32 cores (AMD Rome 7H12) and 64\,GB of RAM.

The maximum number of grid cells used for the simulations presented in the main text were $0.5\times10^7$ for single streamers and $1.9\times10^7$ for branching streamers. The minimal grid size in simulations was $12\,\mu\mathrm{m}$.

\section*{References}.

\bibliographystyle{unsrt}
\bibliography{branching-ref}

\end{document}